\documentclass[11pt]{article}
\usepackage{moriond,epsfig}
\usepackage{amssymb}

\def\be{\begin{equation}}
\def\ee{\end{equation}}
\def\bea{\begin{eqnarray}}
\def\eea{\end{eqnarray}}

\begin{document}
\newcommand{\gm}{\gamma}
\newcommand{\minute}{\mathrm{min}}
\newcommand{\gev}{\mathrm{GeV}}
\newcommand{\tev}{\mathrm{TeV}}
\newcommand{\etev}{E_{\mathrm{TeV}}}
\newcommand{\dnde}{\frac{\mathrm{d}\phi}{\mathrm{d}E}}
\newcommand{\unitphi}{\mathrm{cm}^{-2}\:\mathrm{s}^{-1}}
\newcommand{\unitdnde}{\mathrm{cm}^{-2}\:\mathrm{s}^{-1}\:\mathrm{TeV}^{-1}}
\newcommand{\phiint}{\Phi_{>250\:\gev}}
\vspace*{4cm}
\title{VERY HIGH-ENERGY GAMMA-RAY SOURCES\\AS SEEN BY THE CAT IMAGING TELESCOPE}

\author{F. PIRON, for the CAT collaboration}

\address{Groupe d'Astroparticules de Montpellier, CC 085 - B\^at. 11, Universit\'e de Montpellier II,\\
 Place Eug\`ene Bataillon, 34095 Montpellier Cedex 5 -- {\tt piron@in2p3.fr}}

\maketitle\abstracts{
To date, only three objects have been firmly established as very high-energy gamma-ray sources in the Northern sky:
the Crab nebula, which is a plerion, and the two blazars Markarian~501 and Markarian~421.
This paper reviews the most striking results obtained for these sources by the CAT atmospheric Cherenkov imaging telescope.
}

\section{Introduction}
The CAT (Cerenkov Array at Th\'emis) $\gm$-ray detector~\cite{Barrau98} operates above $250\:\gev$ since Autumn 1996. Its
observation program has been largely devoted to the study of blazars.
These objects are radio-loud active galactic nuclei (AGN's) with a relativistic jet which is pointed directly to the observer.
The jet emission dominates that of the accretion disk (and of the host galaxy) over a large energy domain.
Thus, blazar observations offer the possibility of investigating the physics of jets more deeply, including particle
acceleration and energy extraction in the vicinity of the AGN's central ``engine''.

Until now, however, only two blazars have been clearly detected~\footnote{See the discussion in Weekes (1999).} at very
high-energies (VHE) from the Northern hemisphere, these are Markarian~501 (Mkn~501) and Markarian~421 (Mkn~421).
Since both are variable sources, their study by an atmospheric Cherenkov
experiment requires a kind of ``test beam'' to check that the detector response is well under control. This is
further mandatory if one wishes to compare the temporal and spectral properties of these blazars between different
observation epochs. In VHE $\gm$-ray astronomy, the Crab nebula is a strong source with a steady flux which can be used as
a standard candle.
In the following, we review briefly the results obtained by CAT on this source over four years, before
dealing with the observations of Mkn~501 and Mkn~421.

\section{Non-variability of the Crab nebula VHE emission}
\begin{center}
\begin{figure}[h]
\vspace*{-.5cm}
\hbox{
(a)
\hspace*{-1.cm}
\epsfig{file=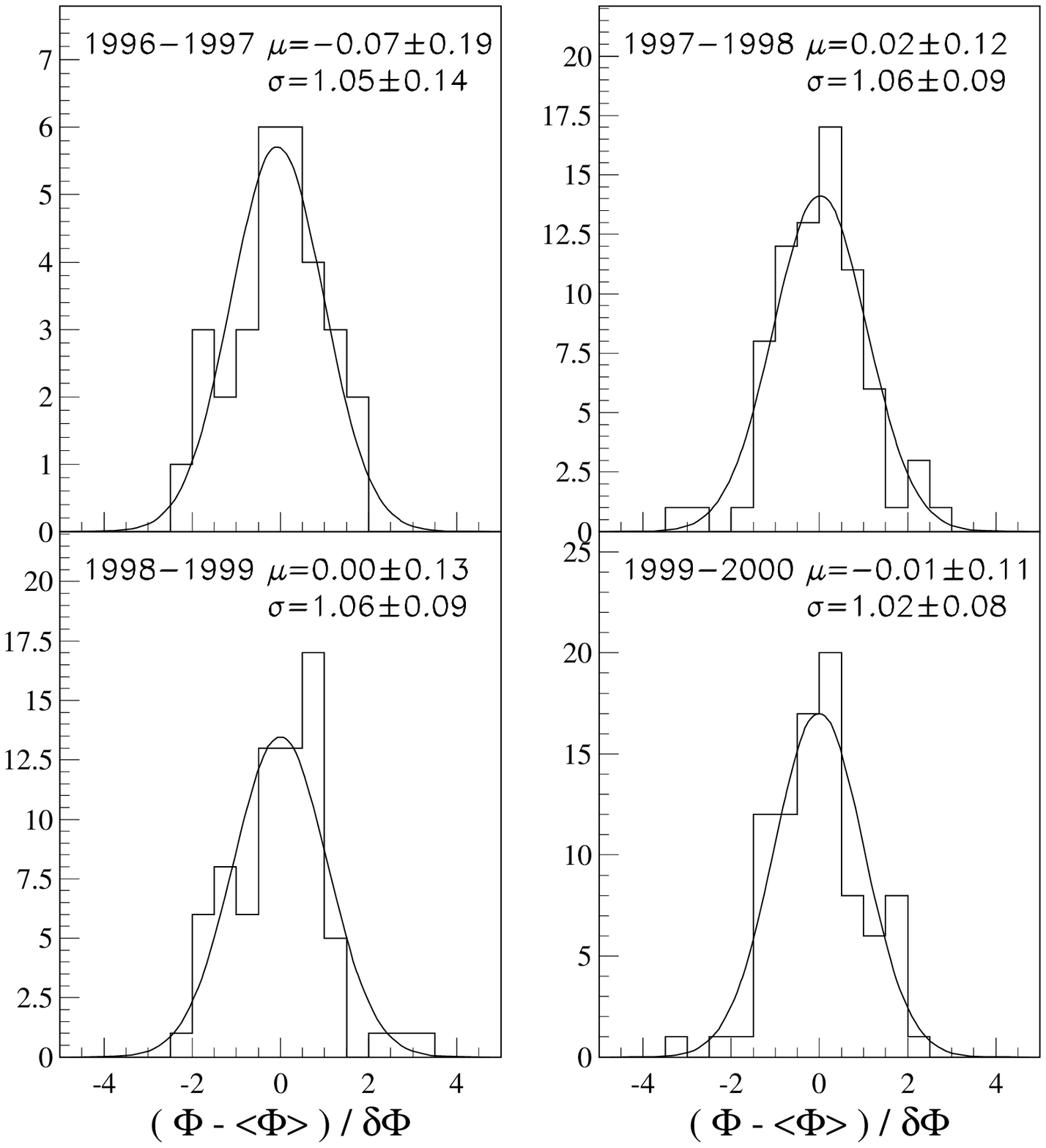,width=0.45\linewidth,clip=
,bbllx=60pt,bblly=0pt,bburx=520pt,bbury=520pt}
\hspace*{.2cm}
\vbox{
\hbox{
\hspace*{.5cm}
(b)
\hspace*{-1.cm}
\epsfig{file=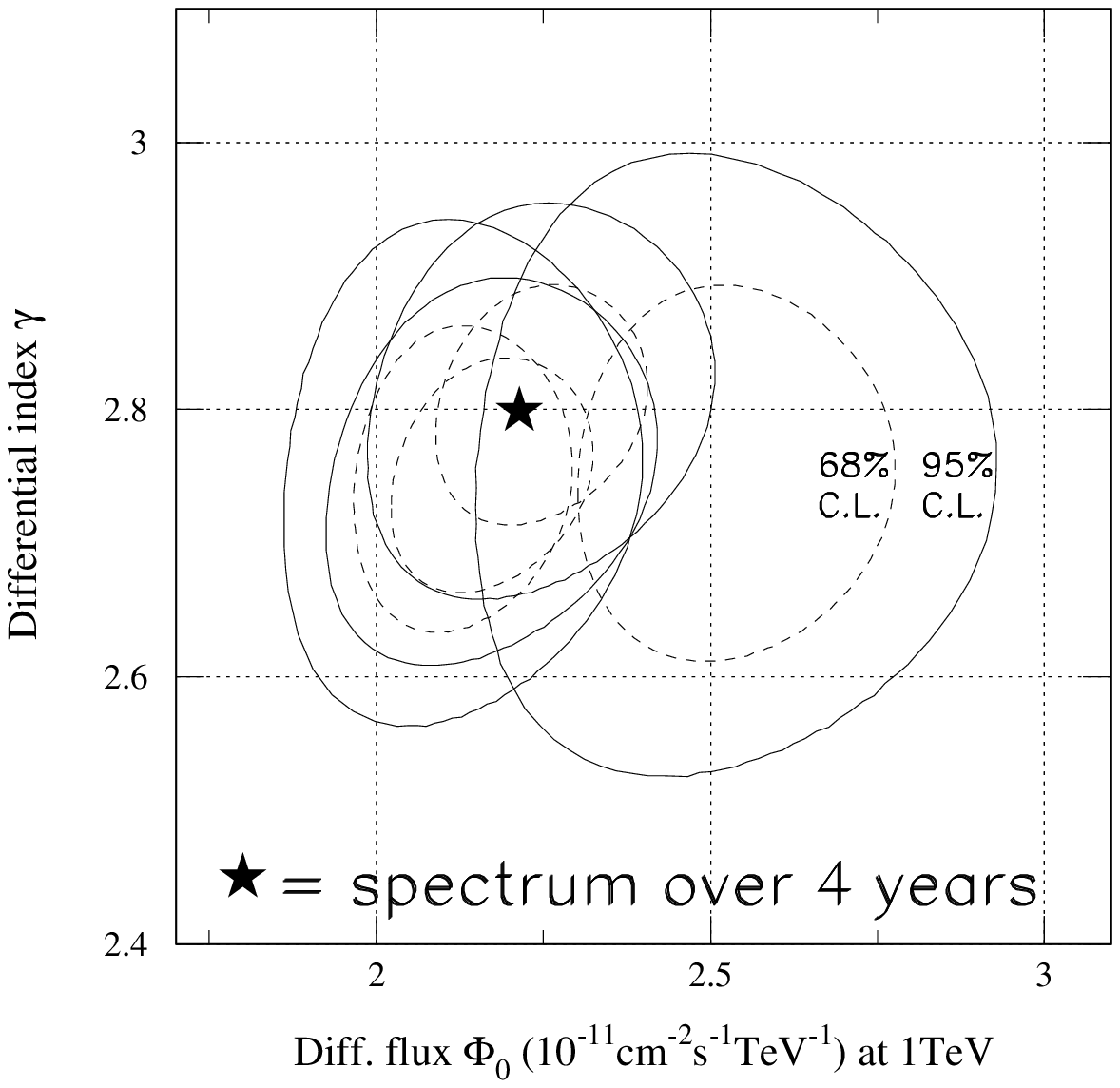,width=0.39\linewidth,clip=
,bbllx=30pt,bblly=40pt,bburx=380pt,bbury=380pt}
}
\vspace*{0.15cm}
\parbox[h]{0.5\linewidth}{
\caption[]
{Stability of the Crab nebula flux and spectrum measured by CAT. For each year between 1996 and 2000 we give:
(a) the flux residuals distribution;
(b) the confidence level contours in the plane $\{\phi_0,\;\gm\}$ of spectral parameters.
}
\label{FigStab}
}
\vspace*{0cm}
}
}
\end{figure}
\end{center}
\vspace*{-.5cm}
The study of the Crab nebula VHE flux, as recorded by CAT between 1996 and 2000, is detailed in Piron (2000).
The time-averaged integral fluxes above $250\:\gev$ for the four years are
$14.9\pm0.9$, $13.3\pm0.6$, $14.5\pm0.7$ and $14.2\pm0.7\times10^{-11}\:\unitphi$, with a
mean value $\Phi^\mathrm{CN}$$=$$14.10\pm0.35\times10^{-11}\:\unitphi$. As can be seen in Fig.~\ref{FigStab}(a), the
flux residuals within each year obey a Gaussian distribution with mean value $\mu$$\simeq$$0$ and variance
$\sigma$$\simeq$$1$. This shows that the dispersion of all flux measurements is here purely statistical, and
confirms the stability of the source emission (and of the detector !) down
to a short time-scale, i.e. that of a single data acquisition ($\sim$$30\:\minute$).

The spectral analysis of CAT data is based on a likelihood method~\cite{PironThese,Piron01}.
It assumes a given parameterization for the spectral shape, and two simple hypotheses are successively
considered for the differential $\gm$-ray spectrum: a power law, $\phi_0\:\etev^{-\gm}$, and a curved shape.
When applied to the $\sim$$100\:$hours of data taken on the Crab nebula, this method indicates the
absence of any curvature between $0.5$ and $13.0\:\tev$, and leads to the following power-law spectrum: 
$\displaystyle\dnde=(2.21\pm0.05^\mathrm{stat}\pm0.60^\mathrm{syst})10^{-11}
\etev^{-2.80\pm0.03^\mathrm{stat}\pm0.06^\mathrm{syst}}\:\unitdnde$.
The results are very stable from one year to an other, as illustrated in Fig.~\ref{FigStab}(b) which shows the
confidence level contours in the plane $\{\phi_0,\;\gm\}$ of spectral parameters. Finally, all measurements shown in
Fig.~\ref{FigStab} make us very confident in the analysis of variable sources like Mkn~501 and Mkn~421.
 
\section{VHE temporal and spectral properties of Mkn~501 and Mkn~421}
\begin{center}
\begin{figure}[t]
\hbox{
(a)
\hspace*{-1.cm}
\epsfig{file=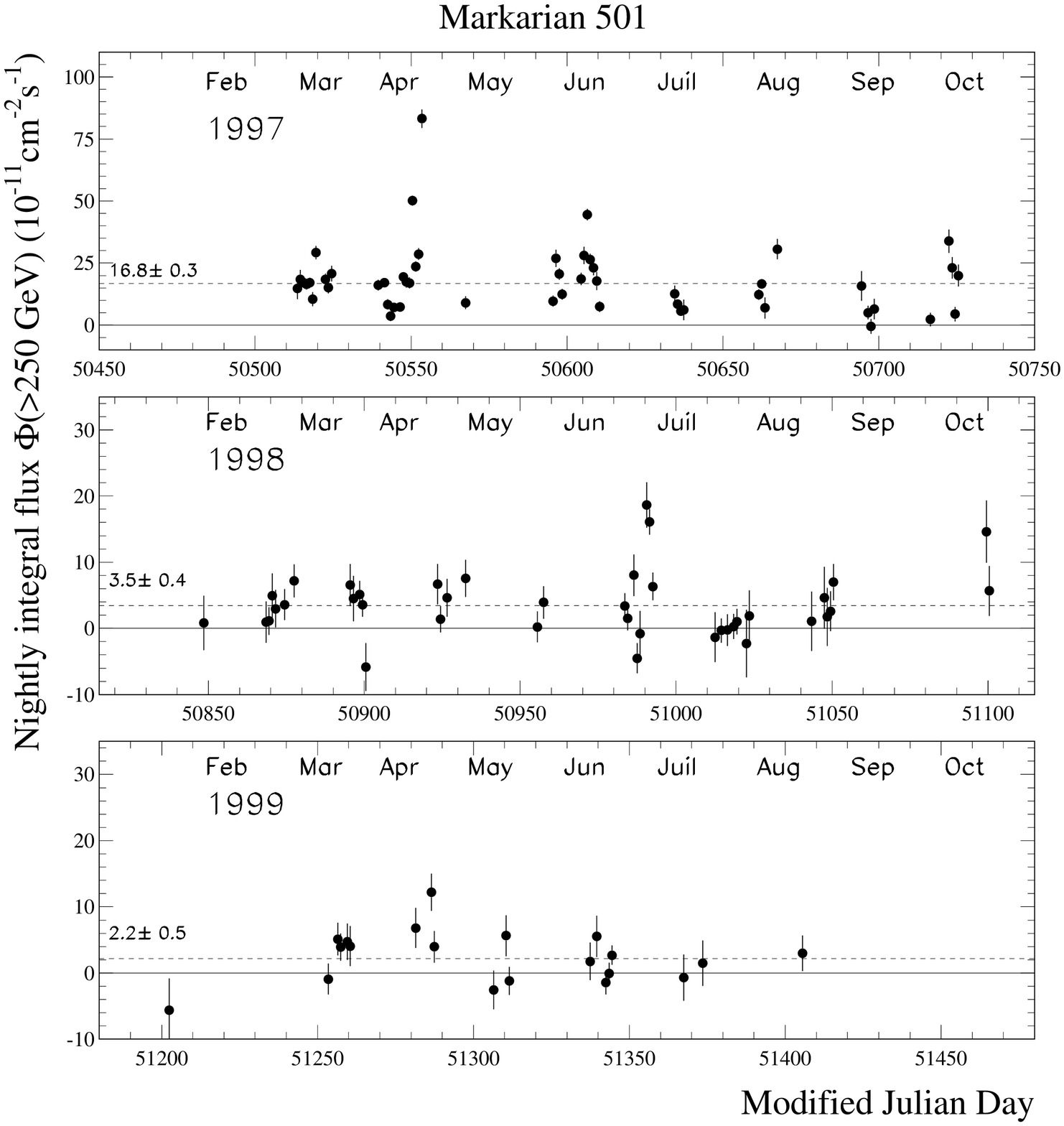,width=0.49\linewidth,clip=
,bbllx=10pt,bblly=0pt,bburx=680pt,bbury=690pt}
\hspace*{.2cm}
(b)
\hspace*{-1.cm}
\epsfig{file=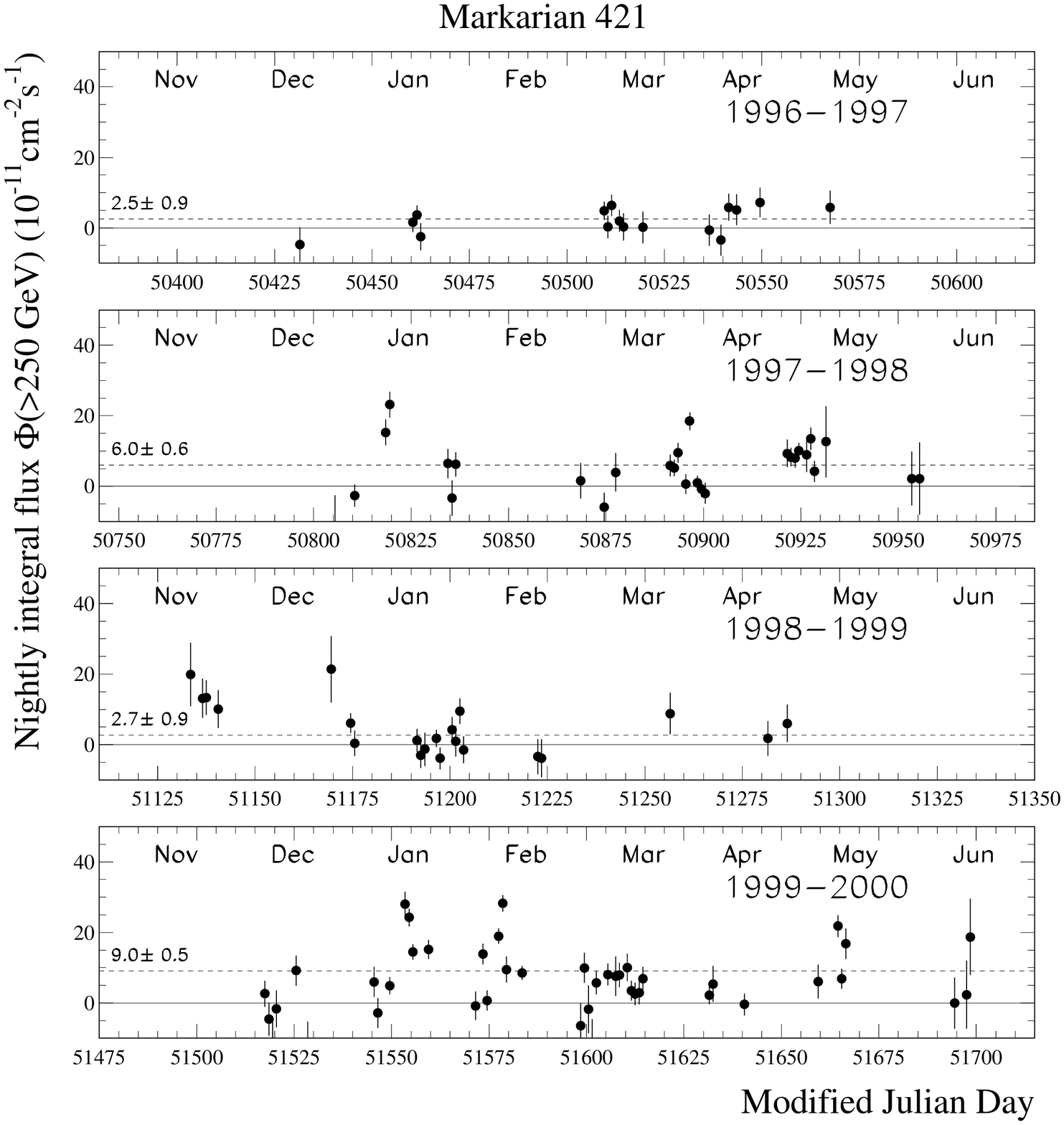,width=0.49\linewidth,clip=
,bbllx=10pt,bblly=0pt,bburx=680pt,bbury=690pt}
}
\caption[]
{Mkn~501 (a) and Mkn~421 (b) nightly-averaged integral flux above $250\:\gev$ as recorded by CAT between 1996 and 2000.
Dashed lines show the mean flux for each observation period.
}
\label{FigCL}
\end{figure}
\end{center}

\begin{center}
\begin{figure}[t]
\vbox{
\hbox{
\epsfig{file=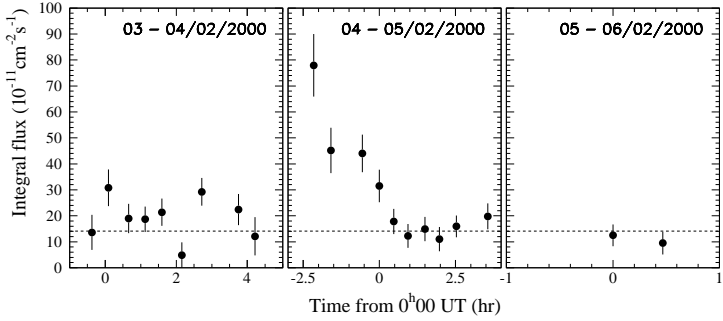,width=0.6\linewidth,clip=
,bbllx=10pt,bblly=25pt,bburx=800pt,bbury=385pt}
\hspace*{0.01\linewidth}
\parbox[h]{0.39\linewidth}{
\vspace*{-5cm}
\caption[]{Mkn~421 integral flux above $250\:\gev$, as seen by CAT between 3 and 6 February, 2000 (MJD 51577 to 51580).
Each point stands for a $\sim$$30\:\minute$ data acquisition and the dashed lines show the Crab nebula level emission.
During the second night, the flux of Mkn~421 decreased by a factor of $2$ in the first hour of observation.
}
\label{FigVar}
}}
}
\end{figure}
\end{center}

The observations of these two blazars by CAT are detailed in the literature~\cite{Djannati99,Piron01}.
Their light curves, as sampled by CAT, are shown in Fig.~\ref{FigCL}(a) and (b), respectively.
Both sources are highly variable, with long periods of intense VHE activity in which a few series of bursts can be
distinguished. This is particularly the case of Mkn~501 in 1997 and of Mkn~421 in 1998 and 2000.
During these flaring periods both sources underwent a large night-to-night variability on several occasions.
However, Mkn~421 is the only source which can also exhibit important flux variations within a single night.
For instance, Mkn~421 light curves for three nights from the 3$^\mathrm{rd}$ to the 5$^\mathrm{th}$ February
are shown in Fig.~\ref{FigVar}. While the fluxes recorded by CAT
during the first and last nights were stable, respectively $\phiint$$\simeq$$1.3\:\Phi^\mathrm{CN}$ and
$\phiint$$\simeq$$0.7\:\Phi^\mathrm{CN}$, the source activity
changed dramatically in a few hours during the second night. The CAT telescope started observation after
the flare maximum while the source flux was at a level of $5.5\:\Phi^\mathrm{CN}$. This is comparable to
the highest $\tev$
flux ever recorded, i.e., that of Mkn~501 during the night of April 16$^\mathrm{th}$, 1997 (see Fig.~\ref{FigCL}(a)).
After this first episode, Mkn~421 intensity was reduced by a factor of $2$ in $1$~hour and by a factor of $5.5$ in
$3$~hours. A simple causality argument implies that the $\gm$-ray emitting region
must be very compact here, with a size $\lesssim$$10$~light-hours if one assumes a typical value of $10$ for the
geometric Doppler factor (which reduces the time-scale in the observer frame).
\vspace*{-1.2cm}

\begin{center}
\begin{figure}[t]
\hbox{
(a)
\hspace*{-1.cm}
\epsfig{file=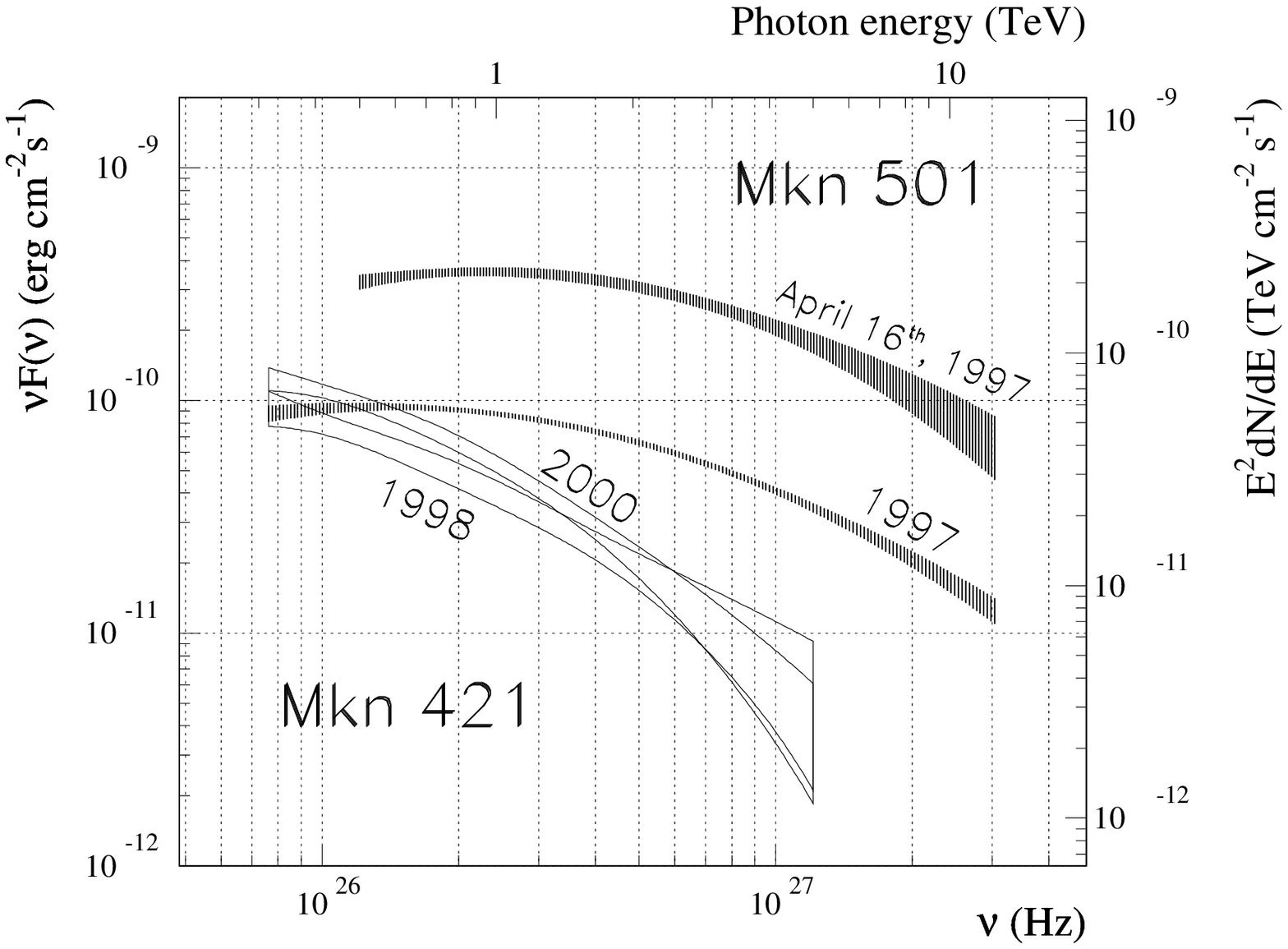,width=0.5\linewidth,clip=
,bbllx=25pt,bblly=28pt,bburx=550pt,bbury=440pt}
\hspace*{.2cm}
(b)
\hspace*{-1.cm}
\epsfig{file=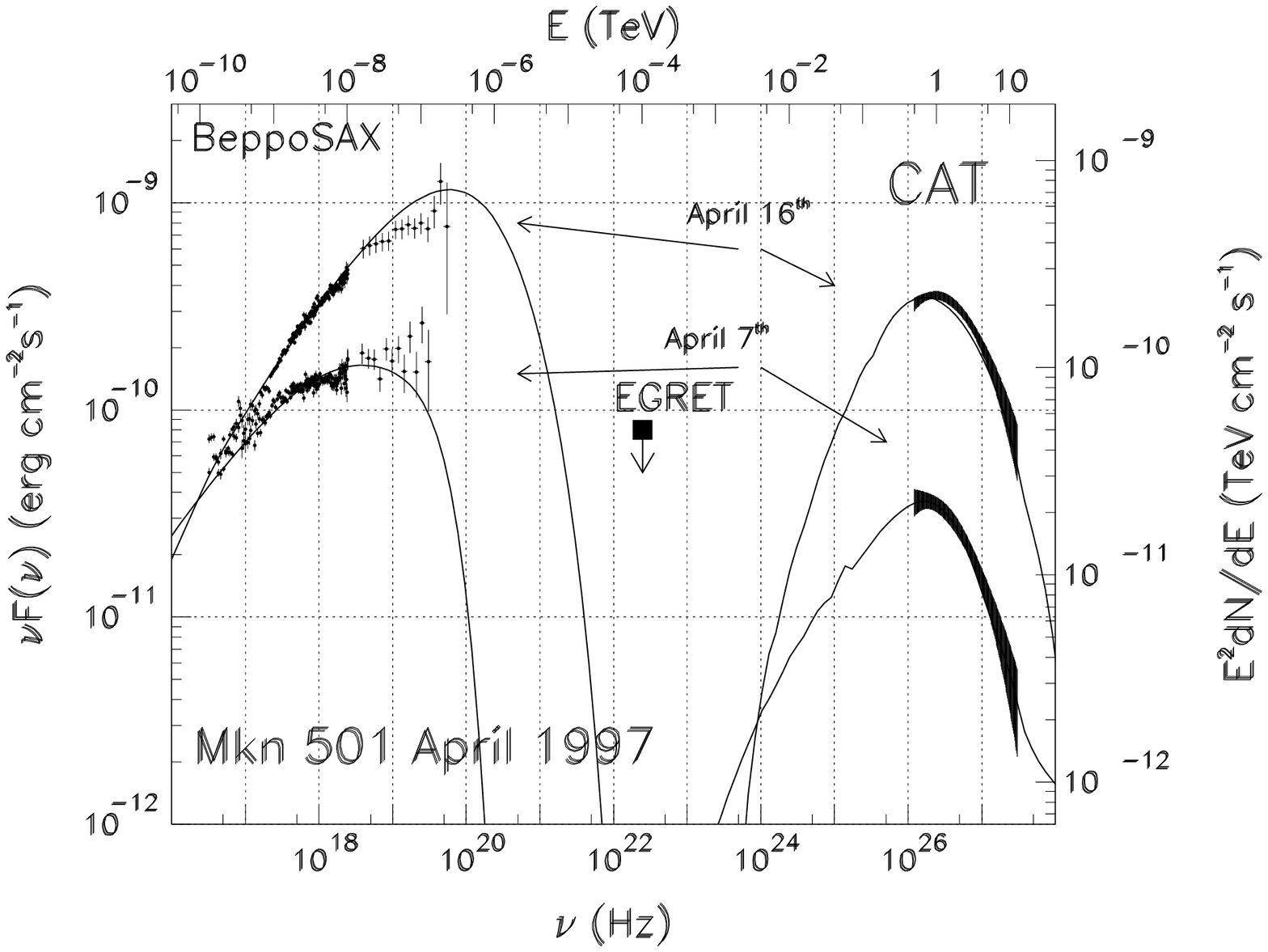,width=0.5\linewidth,clip=
,bbllx=0pt,bblly=0pt,bburx=567pt,bbury=425pt}
}
\caption[]{
Spectral energy distributions (SEDs) of Mkn~501 and Mkn~421. VHE spectra are represented by an area showing the $68$\%
confidence level contour given by the likelihood method with the assumption of a curved spectrum.
(a) VHE SEDs as measured by CAT; (b) Mkn~501 X-ray and VHE SEDs for April 7$^\mathrm{th}$ and 16$^\mathrm{th}$, 1997, as
simultaneously measured by Beppo-SAX~\cite{Pian} and CAT~\cite{Djannati99,PironThese}.
Full lines come from a homogeneous SSC model~\cite{PironThese}.
}
\label{FigSp}
\end{figure}
\end{center}

Fig.~\ref{FigSp}(a) shows the spectral energy distributions (SEDs) of Mkn~501 and Mkn~421. Mkn~501 showed a clearly curved
spectrum in 1997, with a $\gm$-ray peak lying above the CAT threshold. Mkn~421 is less extreme since its spectra do not show
any significant curvature~\footnote{As can be seen in
Fig.~\ref{FigSp}(a), there is in fact some indication of curvature for the 2000 time-averaged spectrum of Mkn~421. This result
is further discussed in Piron et al. (2001).}.
In the framework of leptonic models (Tavecchio, these proceedings), which succesfully explain the SED of Mkn~501 in the
X-ray and VHE $\gamma$-ray ranges (see also our fit in Fig.~\ref{FigSp}(b), which is based on a homogeneous Synchrotron
Self-Compton model~\cite{PironThese}), these results imply that the peak energy of the inverse Compton contribution of
Mkn~421 SED is significantly lower than the C{\small AT} detection threshold.
This is not surprising since the corresponding synchrotron peak is known to be lower than that of Mkn~501, and
since leptonic models predict a strong correlation between X-rays and $\gamma$-rays. 

\vspace*{-1.15cm}
\section{Conclusion}
Although the observations of Mkn~501 and Mkn~421 are quite numerous,
alternative scenarios~\cite{Rachen99,Muecke01} other than leptonic models are still successfull in interpreting their SEDs.
In the future the study of the dynamic aspects of blazar jet emission, including the temporal and spectral correlations between
various wavelengths, is thus required in order to accurately constrain existing models, and to understand the particle
acceleration and cooling processes occuring at the sub-parsec scale in jets.
It should help to discriminate between these models and allow, in particular,
to address more deeply the crucial problem of the plasma jet content.

For instance, Beppo-SAX and CAT have observed a correlated X-ray and $\gm$-ray spectral hardening during the flares of Mkn~501 in 1997
(see Djannati-Ata\"{\i} et al. (1999)). This behaviour is illustrated in Fig.~\ref{FigSp}(b) by the shift of the entire SED between
April 7$^\mathrm{th}$ and 16$^\mathrm{th}$.
A simple correlation has been observed many times for Mkn~421~\cite{Maraschi99,Takahashi99}, but always
in terms of integrated (and not differential) fluxes due to the lack of statistics. Moreover, the VHE spectral variability
of Mkn~421 has not been clearly proven yet (we discuss this problem in Piron et al. (2001)). 
Hopefully, the recent huge bursts recorded from this source in the beginning of 2001 should help to clarify this point.

\end{document}